# Spin diffusion length of Permalloy using spin absorption in lateral spin valves

E. Sagasta[1], Y. Omori[2], M. Isasa[1], Y. Otani[2,3], L. E. Hueso[1,4] and F. Casanova[1,4]

[1]*CIC nanoGUNE, 20018 Donostia-San Sebastian, Basque Country, Spain*

[2]*Institute for Solid State Physics, University of Tokyo, Kashiwa, Chiba 277-8581, Japan*

[3]*RIKEN-CEMS, 2-1 Hirosawa, Wako, Saitama 351-0198, Japan*

[4]*IKERBASQUE, Basque Foundation for Science, 48011 Bilbao, Basque Country, Spain*

**Abstract**

We employ the spin absorption technique in lateral spin valves to extract the spin diffusion length of Permalloy (Py) as a function of temperature and resistivity. A linear dependence of the spin diffusion length with conductivity of Py is observed, evidencing that Elliott-Yafet is the dominant spin relaxation mechanism in Permalloy. Completing the data set with additional data found in literature, we obtain $\lambda_{Py} = (0.91\pm0.04)$ (f$\Omega$m$^2$) /$\rho_{Py}$.

Spintronics is the field of electronics aiming at exploiting, apart from the charge, the spin degree of freedom of electrons, whereas in conventional electronics only the charge of the electron is employed. There is a significant difference between charge and spin currents: the first one is a conservative magnitude, whereas the second one is not[1]. Conduction electrons find different ways to relax their spin, *i.e.*, to bring an unbalanced population of spin states into equilibrium. This relaxation occurs due to the spin-orbit coupling (SOC)[2,3].

There are several spin relaxation mechanisms based on the SOC. Elliott-Yafet is the governing mechanism in metals with inversion symmetry in the absence of magnetic impurities[4,5]. The spin of the electron in these elements interacts with the local electric field generated by the lattice ions (phonons), non-periodic impurities or crystal grain boundaries, resulting in a spin-flip scattering. In this case, spin relaxation time $\tau_s$ is proportional to the momentum relaxation time $\tau$, $\tau_s \propto \tau$, as more frequent momentum scattering means more frequent spin flipping. In contrast, spin relaxation due to D´yakonov-Perel mechanism arises in systems with lack of spatial inversion symmetry[6]. The spin-up and spin-down energy levels in the conduction bands are split, generating a momentum-dependent effective magnetic field which leads to spin precession and, hence, spin relaxation. The smaller the momentum relaxation time is, the less time the spin has to change its direction by precession around the magnetic field, making the spin relaxation time longer. Then, $\tau_s \propto 1/\tau$.

It is well known that in light metals with weak SOC, such as Cu[7-9], Ag[10-12] and Al[8], Elliott-Yafet mechanism is dominating the spin relaxation[13]. In heavier elements that have strong SOC, such as Pt and Ta, the crystallinity and the thickness of the layers influence significantly in the spin relaxation mechanism, and both mechanisms have recently been identified in Pt and Ta[14-17]. The origin of spin-flip scattering in ferromagnetic elements has been barely studied, although these materials play significant role in the field of spintronics and, thus, it is fundamental to know the mechanisms that contribute to the spin relaxation. A theoretical model presented by L. Berger[18] extends the Elliott theory of spin relaxation in metals and semiconductors to metallic ferromagnets. However, no experimental work has been performed so far to address this purpose.

In this work we study the spin relaxation mechanism in Permalloy (Py, $Ni_{80}Fe_{20}$). This ferromagnetic metal is widely employed in spintronics for spin current injection and detection in many techniques, such as electrical spin injection and detection[19], spin pumping[20], spin transfer torque[21] and thermal spin injection[22,23]. We employ the spin absorption technique[16,24-29] in lateral spin valves to extract the spin diffusion length of Py, $\lambda_{Py}$, as a function of temperature and resistivity, $\rho_{Py}$. The spin diffusion length, defined as $\lambda = \sqrt{D\tau_s}$ with $D$ being diffusion constant, is the mean distance over which the electrons diffuse between spin-flip scattering events[30]. This quantification of $\lambda_{Py}$ as a function of temperature and resistivity is lacking in literature and assumptions of its tendency were necessary to realize in order to quantify other spin dependent parameters. We find that the obtained results for $\lambda_{Py}$, together with the ones reported in literature, show a linear dependence with $1/\rho_{Py}$ which clearly indicates that Elliott-Yafet is the main spin relaxation mechanism in Py.

To this end, we fabricate a sample (Sample 1) with two types of devices (see a scanning electron microscopy image of Sample 1 in Fig. 1(a)). The first type of device consists of a Py/Cu lateral spin valve (LSV), where the Py injector and Py detector are connected by a Cu channel with the Py interelectrode distance $L$. The second type of device is a Py/Cu LSV that contains an additional Py nanowire in between the Py injector and detector. The comparison of the non-local signals obtained in each of the devices allows us to study the spin relaxation in the middle Py wire. The fabrication consists of two step e-beam lithography, metal deposition and lift-off. First, Py wires are patterned and 32 nm of Py are e-beam evaporated in ultra-high vacuum ($3.3\times10^{-8}$ mbar) at 0.8 Å/s. Then, the Cu channel is patterned and 95 nm of Cu are thermally evaporated in ultra-high vacuum ($1.8\times10^{-8}$ mbar) at 1.5 Å/s. Before the Cu deposition, an ion-milling process is performed, in order to obtain good interface quality between Py and Cu[7]. All non-local transport measurements were carried out in a liquid-He cryostat, applying an external magnetic field $H$ and varying temperature $T$, using a "dc reversal" technique[31].

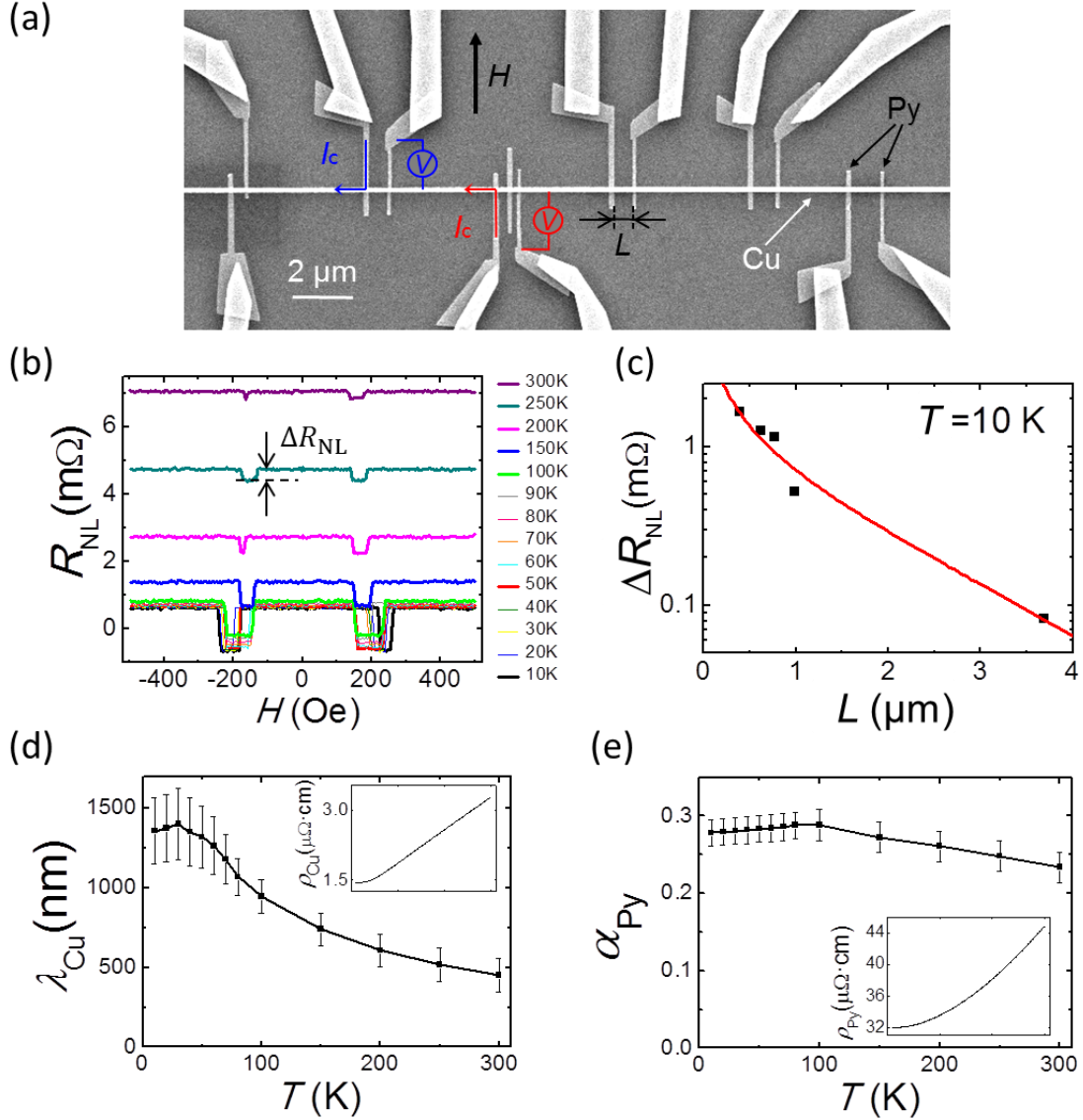

FIG. 1. (a) SEM of Sample 1 containing six Py/Cu LSVs with different interelectrode distances $L$. One of them, the third LSV from the left side, has an additional Py wire in between the Py electrodes where the spin absorption will occur. The non-local measurement configuration, the materials (Py, Cu) and the direction of the applied magnetic field are shown. (b) Non-local resistance as a function of magnetic field measured at different temperatures in the device where $L$=650 nm using the configuration shown in a) and applying $I_c$=100μA. The spin signal ($\Delta R_{NL}$) has been tagged. (c) $\Delta R_{NL}$ as a function of $L$ at 10 K. Red solid line is the fitting of the experimental data, represented by black squares, to Eq. (1) from which we extract (d) the spin diffusion length of Cu, $\lambda_{Cu}$, and (e) the spin polarization of Py, $\alpha_{Py}$, as a function of the temperature. Insets in (d) and (e) correspond to the temperature dependence of the resistivity of Cu and Py, respectively. The scale in the horizontal axis of the insets is the same as in their respective main panel.

When a spin-polarized current is injected from one Py electrode into the Cu channel, a spin accumulation is created at the Py/Cu interface which diffuses along both sides of the Cu channel, creating a pure spin current that is detected as a voltage by the second Py electrode. Normalizing the measured voltage to the injected current $I_C$, the non-local resistance $R_{NL}$ is defined. This value changes sign when the relative magnetization of the two Py electrodes is switched from parallel to antiparallel configuration by sweeping $H$. The change from positive to negative $R_{NL}$ is defined as the spin signal

$\Delta R_{NL}$. The one-dimensional spin-diffusion model for transparent interfaces gives the following expression for the spin signal[32-35]:

$$\Delta R_{NL} = \frac{4\alpha_{Py}^2 R_{Cu}}{\left[2+\frac{R_{Cu}}{R_{Py}}\right]^2 e^{\frac{L}{\lambda_{Cu}}} - \left[\frac{R_{Cu}}{R_{Py}}\right]^2 e^{-\left(\frac{L}{\lambda_{Cu}}\right)}} \quad (1)$$

where $R_{Cu}=\frac{\lambda_{Cu}\rho_{Cu}}{w_{Cu}t_{Cu}}$ and $R_{Py}=\frac{\lambda_{Py}\rho_{Py}}{w_{Py}w_{Cu}(1-\alpha_{Py}^2)}$ are the spin resistances of the Cu channel and Py electrodes, respectively. $\lambda_{Cu}$ and $\lambda_{Py}$ are the spin diffusion length of Cu and Py. $\rho_{Cu}$ and $\rho_{Py}$ are the resistivity of the Cu and Py wires. $w_{Cu}$ and $t_{Cu}$ are the width and thickness of the Cu channel, and $w_{Py}$ the width of the Py electrodes. $\alpha_{Py}$ is the spin polarization of Py.

We measure $R_{NL}$ as a function of $H$ in the devices without the middle Py wire with different interelectrode distance $L$ and at different temperatures (see Fig. 1(b) for $L=$ 650 nm). Figure 1(c) shows the obtained $\Delta R_{NL}$ as a function of $L$ at 10 K. From the fitting of the data to Eq. (1) (red solid line in Fig. 1(c)), we extract $\lambda_{Cu}$ and $\alpha_{Py}$, which are plotted as a function of temperature in Fig. 1(d) and Fig. 1(e), respectively. In order to perform the fitting, we measure experimentally all the dimensions and resistivities of the wires that form the device. The resistivities of Py and Cu wires are plotted as a function of temperature in the insets of Fig. 1(d) and Fig. 1(e), respectively. $\lambda_{Py}$ was first assumed to be 5 nm at 10 K and considered a temperature dependence coming from the resistivity in the form $\lambda_{Py}=$const$/\rho_{Py}$ (Ref. 7).

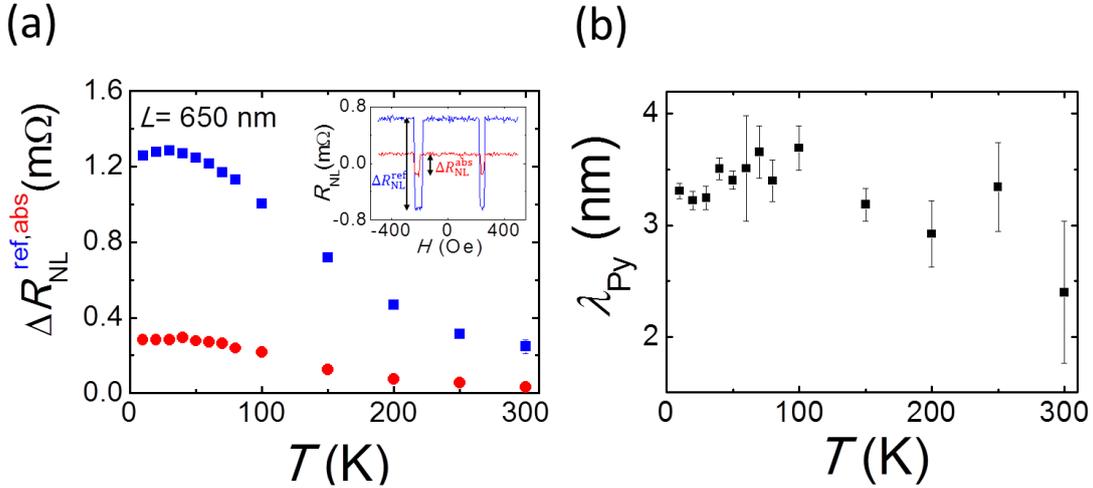

FIG. 2. (a) Spin signal as a function of temperature for the reference Py/Cu LSV (blue squares) and for Py/Cu LSV, both in Sample 1, with a middle Py wire (red circles) using $I_c=100\mu A$. The distance between the injector and detector is the same in both devices. Inset: non-local resistance as a function of the magnetic field at 10 K for the reference Py/Cu LSV (blue) and the Py/Cu LSV with a middle Py wire (red). The reference spin signal ($\Delta R_{NL}^{ref}$) and the spin signal with Py absorption ($\Delta R_{NL}^{abs}$) are tagged. (b) Spin diffusion length of Py as a function of the temperature obtained from the data in a) using Eq. (2).

Next, we measure the non-local resistance in the LSV with the middle Py wire. The inset in Fig. 2(a) shows $R_{NL}$ as a function of $H$ for the reference LSV (blue line) and the LSV with the middle Py wire (red line) measured in the configuration shown in Fig 1(a). The distance between Py injector and detector in both LSVs is 650 nm. In the latter case, the middle Py wire absorbs part of the spins that are flowing in the Cu channel, reducing the spin signal. The spin signal obtained for each type of LSV at different temperatures is shown in Fig. 2(a). The ratio of both spin signals, obtained from the one-dimensional spin-diffusion model for transparent interfaces, is given by[24,27]:

$$\eta = \frac{\Delta R_{NL}^{abs}}{\Delta R_{NL}^{ref}} = \frac{2Q_M \left[\sinh\left(\frac{L}{\lambda_{Cu}}\right) + 2Q_{Py}e^{\left(\frac{L}{\lambda_{Cu}}\right)} + 2Q_{Py}^2 e^{\left(\frac{L}{\lambda_{Cu}}\right)}\right]}{\left[\cosh\left(\frac{L}{\lambda_{Cu}}\right) - 1\right] + 2Q_M \sinh\left(\frac{L}{\lambda_{Cu}}\right) + 2Q_{Py}\left[e^{\left(\frac{L}{\lambda_{Cu}}\right)}(1+Q_{Py})(1+2Q_M) - 1\right]} \quad (2)$$

where $Q_{Py}=R_{Py}/R_{Cu}$ and $Q_M=R_M/R_{Cu}$, being $R_M$ the spin resistance of the middle wire. In our case, as $R_M=R_{Py}$, $Q_M=Q_{Py}$. From this equation, the value of $\lambda_{Py}$ can be obtained.

Figure 2(b) shows $\lambda_{Py}$ extracted from Eq. (2) as a function of temperature. The obtained $\lambda_{Py}$ is different from the one originally assumed. With the new $\lambda_{Py}$, we can make another iteration with Eq. (1) and Eq. (2) to recalculate $\lambda_{Cu}$, $\alpha_{Py}$ and $\lambda_{Py}$. Iterations were performed until $\lambda_{Cu}$, $\alpha_{Py}$ and $\lambda_{Py}$ parameters converged in a self-consistent manner. Results obtained in the second cycle are shown by blue solid lines in Fig. 3. Although the parameter $\lambda_{Cu}$ barely changes from the first to the second cycle, $\alpha_{Py}$ varies quite significantly. In the third cycle, the convergence is attained for the three parameters, see red solid line in Fig. 3. The obtained $\lambda_{Cu}$, $\alpha_{Py}$ and $\lambda_{Py}$ values are consistent with the values that are reported in literature, see Tables I and II.

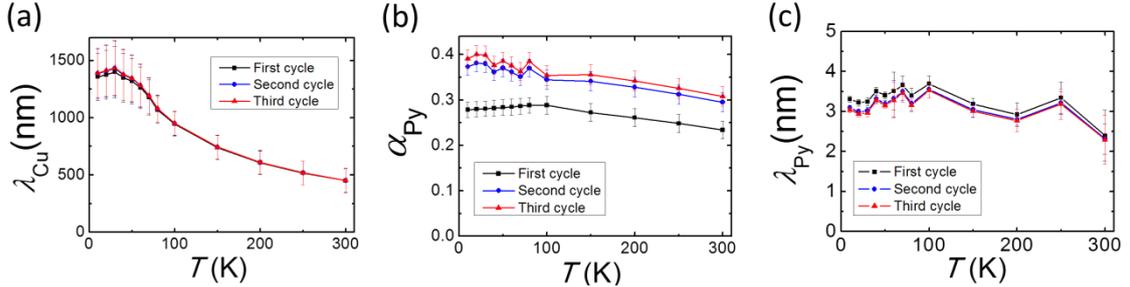

FIG. 3. Results of three self-consistent cycles for (a) spin diffusion length of Cu, (b) spin polarization of Py and (c) spin diffusion length of Py, as a function of temperature. First, second and third cycles are represented by black, blue and red solid lines, respectively. Data corresponds to Sample 1.

We fabricated an additional Py/Cu LSV (Sample 2) with a thinner middle Py wire (9 nm) than in the previous one (Sample 1), in order to increase its resistivity. The dimensions and characteristics of the Cu channel and Py injector and detector in this new sample are the same as in the previous one. We measured the spin signal from 10 K to 50 K and extracted the spin diffusion length by employing Eq. (2). The obtained results have been added in Table II.

TABLE I. Spin diffusion length of Cu and spin polarization of Py extracted from literature and this work. Temperature and resistivity of Cu are included.

| T (K) | $\rho_{Cu}$ (μΩcm) | $\lambda_{Cu}$ (nm) | $\alpha_{Py}$ | Ref. |
|---|---|---|---|---|
| 10 | 0.69 | 1000 | 0.58 | 9 |
| 10 | 1.26 | 1020 | 0.40 | 7 |
| 10 | 1.2 | 770 | 0.39 | 36 |
| 10 | 1.44 | 1390±200 | 0.39±0.02 | Sample 1 |
| 80 | 1.2 | 1300 | 0.35 | 37 |

| 250 | 2.4 | 380 | 0.34 | 36 |
| 300 | 2.35 | 380 | 0.49 | 9 |
| 300 | 2.08 | 500 | 0.25 | 34 |
| 300 | 2.90 | 410 | 0.34 | 7 |
| 300 | 3.30 | 450±100 | 0.31±0.02 | Sample 1 |

TABLE II. Spin diffusion length and resistivity of Py extracted from literature and this work. Temperature is included.

| $T$ (K) | $\rho_{Py}$ (μΩcm) | $\lambda_{Py}$ (nm) | Ref. |
|---|---|---|---|
| 4.2 | 12 | 5.5±1 | 30,38 |
| 10 | 17.1 | 5 | 9 |
| 10 | 32 | 3.04±0.06 | Sample 1 |
| 10 | 80.2 | 1.4±0.2 | Sample 2 |
| 77 | - | 4.3±1 | 39 |
| 300 | - | 2.5 | 40 |
| 300 | 23.1 | 4.5 | 9 |
| 300 | 26.8 | 3 | 34 |
| 300 | 44 | 2.30±0.61 | Sample 1 |

Figure 4(a) shows that the $\rho_{Py}\lambda_{Py}$ values vary slightly with temperature and are similar for Sample 1 and Sample 2. The obtained values are close to the one given in Ref. 41. Figure 4(b) demonstrates the linear dependence of $\lambda_{Py}$ with the conductivity of Py, $\sigma_{Py}=1/\rho_{Py}$, not only for our samples, but also for the experimental data from the literature. We observe a general linear tendency that fits well to $\lambda_{Py} = (0.91\pm0.04)$ (fΩm$^2$) /$\rho_{Py}$. These plots indicate that the main spin relaxation mechanism in Py is Elliott-Yafet, which is consistent with the theoretical prediction of L. Berger[18].

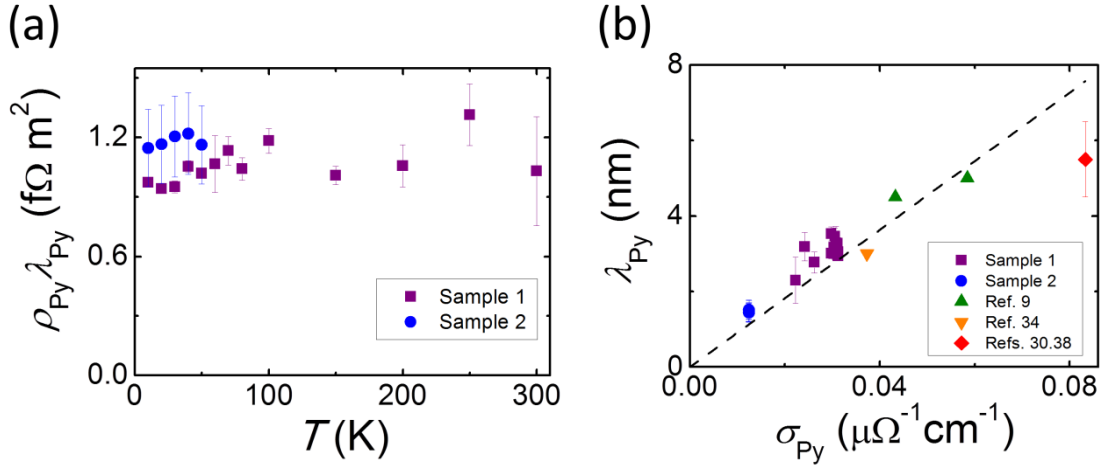

FIG 4. (a) Product of spin diffusion length and resistivity of Py as a function of temperature for Sample 1 and Sample 2. (b) Spin diffusion length of Py as a function of the conductivity. Literature values of $\lambda_{Py}$ are also included for completing the data set. Black dashed line corresponds to the linear fit to all data.

To conclude, we obtained the temperature and resistivity dependence of the spin diffusion length in Py using the spin absorption technique in lateral spin valves. We observe a linear dependence between $\lambda_{Py}$ and $1/\rho_{Py}$ which evidences that Elliott-Yafet is the dominating spin relaxation mechanism in Py.

**Acknowledgements**

We acknowledge support from the Spanish MINECO (Project No. MAT2015-65159-R), from the Regional Council of Gipuzkoa (Project No. 100/16), and from JSPS Grant-in-Aid for Scientific Research on Innovative Area, "Nano Spin Conversion Science" (Grant No. 26103002). E.S. and M.I. thank the Spanish MECD and the Basque Government, respectively, for a Ph.D. fellowship (Grants No. FPU14/03102 and No. BFI-2011-106). Y.O. acknowledges financial support from JSPS through "Research program for Young Scientists" (Grant No. 15J08073) and "Program for Leading Graduate Schools (MERIT)."